\pdfoutput = 1

\documentclass[letterpaper,conference]{IEEEtran}
\IEEEoverridecommandlockouts
\usepackage{graphicx}
\usepackage[utf8]{inputenc} 
\usepackage[T1]{fontenc}
\usepackage{url}
\usepackage{ifthen}
\usepackage{cite}
\usepackage{color}
\usepackage{amsthm}

\ifCLASSINFOpdf
\else
\fi

\usepackage{cite}
\usepackage{graphicx}
\usepackage{hyperref}
\usepackage{amsmath}
\usepackage{amssymb}
\usepackage{bm}
\usepackage{bbm}
\usepackage{algorithmicx}
\usepackage{algorithm}
\usepackage{algpseudocode}
\usepackage{array}
\usepackage{booktabs}
\usepackage{multirow}
\usepackage{makecell}

\begin{document}
%
\title{Finite-Support Capacity-Approaching Distributions for AWGN Channels\thanks{This research is supported in part by National Science Foundation (NSF) grant CCF-1911166. Any opinions, findings, and conclusions or recommendations expressed in this material are those of the author(s) and do not necessarily reflect the views of the NSF}}



%
\author{\IEEEauthorblockN{Derek Xiao\IEEEauthorrefmark{1},
Linfang Wang\IEEEauthorrefmark{1},
Richard D. Wesel\IEEEauthorrefmark{1}}
\IEEEauthorblockA{\IEEEauthorrefmark{1}University of California, Los Angeles, Los Angeles, CA 90095, USA}
Email: derekxiao93@ucla.edu, lfwang@ucla.edu,  wesel@ucla.edu
}


\maketitle

\begin{abstract}
In this paper, the Dynamic-Assignment Blahut-Arimoto (DAB) algorithm identifies finite-support probability mass functions (PMFs) with small cardinality that achieve capacity for amplitude-constrained (AC) Additive White Gaussian Noise (AWGN) Channels, or approach capacity to within less than 1\% for power-constrained (PC) AWGN Channels.  While a continuous Gaussian PDF is well-known to be a theoretical capacity-achieving distribution for the PC-AWGN channel, DAB identifies PMFs with small-cardinality that are, for practical purposes, indistinguishable in performance. We  extend the results of Ozarow and Wyner that require a constellation cardinality of $2^{C+1}$ to approach capacity $C$ to within the shaping loss. PMF's found by DAB approach capacity with {\em essentially no shaping loss} with constellation cardinality of $2^{C+1.2}$. For AC-AWGN channels, DAB characterizes the evolution of minimum-cardinality finite-support capacity-achieving PMFs.
\end{abstract}

%
\IEEEpeerreviewmaketitle

\section{Introduction}


\subsection{Background}
Probabilistic amplitude shaping (PAS) \cite{bocherer_bandwidth_2015,bocherer_achievable_2018,schulte_four_2017,buchali_rate_2016} enables practical coded modulation for any symmetric constellation.  The constellation points do not need to be equally spaced, nor do they need to be equally likely.  PAS rekindles interest in symmetric, low-cardinality probability mass functions (PMFs) that closely approach capacity.

Smith was the first to prove that the capacity-achieving distribution for the amplitude-constraint AWGN (AC-AWGN) channel is unique and has finite-support\cite{smith_information_1971}, and many alternative proofs followed \cite{Chan2005} \cite{Dytso2020}.  Smith also described a computer program capable of finding the AC-AWGN capacity-achieving distributions for any amplitude constraint, fixing noise power. Subsequently, upper bounds on capacity-achieving input cardinality were identified by Yagli et al. \cite{Dytso2020,Yagli2019}, and theoretical conditions for the input cardinality transition SNRs were derived in \cite{Sharma_2008}, as well as approximate SNRs for transitions from binary and ternary optimal signaling to the next higher cardinality. Various upper and lower bounds on the AC-AWGN capacity were identified \cite{Thangaraj_2015,McKellipsa}. 

For the power-constrained AWGN (PC-AWGN) channel, Ungerboeck showed emprically \cite{Ungerboeck_1982} and Ozarow and Wyner proved analytically \cite{Ozarow1990} that one-dimensional constellations with $2^{C+1}$ points could achieve rates within the shaping gain (around 0.25 bits for high SNRs) of the channel capacity. For low SNRs, Wu and Verdu \cite{Wu2010} proved that finite-support PMFs defined by Gauss quadrature achieve information rates that converge exponentially to the PC-AWGN channel capacity. Separately, Huang \cite{Huang2005} finds, specifically for the PC-AWGN channel, that discrete approximations with non-optimized point spacing can nearly reach capacity even when the capacity achieving distribution is known to be continuous.  Mathar and Alirezaei \cite{Mather_2020} argue that discrete distributions can always achieve or closely approach the PC-AWGN capacity. 

The Distributed Allocation Blahut-Arimoto (DAB) algorithm is an iterative algorithm originally used to find the capacity and  capacity-achieving distributions for the binomial channel and molecular channels   \cite{WeselITA2018} \cite{Farsad2020}. 

\subsection{Contributions and Organization}

This paper applies DAB to the problem of identifying capacity-achieving finite-support PMFs for the AC-AWGN channel. Our results match Smith \cite{Smith1971TheIC} and  explore how the capacity achieving distribution evolves with a fixed amplitude constraint as noise power varies.

As its main contribution, this paper applies the DAB algorithm to the problem of finding cardinality constrained input PMFs that can closely approach (within 0.01 bits) the capacity of the PC-AWGN channel with no cardinality constraint. Despite the well-known result that capacity is achieved by a continuous (Gaussian) pdf for the PC-AWGN channel,  our results demonstrate that a finite-support input PMF can always approach the capacity to less than 0.01 bits as long as $\log_2$ of the input cardinality is 1.2 bits above the PC-AWGN capacity, or alternatively the entropy of the input PMF is 0.9 bits above the PC-AWGN capacity.


This paper is organized as follows: Section \ref{sec:DAB-Intro} introduces the DAB approach used for the amplitude constrained channel. Sec. \ref{sec:DAB-AWGN-AC} obtains finite-support PMFs that closely approach the channel capacity for the AC-AWGN channel. Sec. \ref{sec:DAB-AWGN-PC} adapts the DAB approach to PC-AWGN channels, and provides minimum-support PMFs that closely approach the PC-AWGN capacity for a wide range of SNRs.

\section{Dynamic-Assignment Blahut-Arimoto}
\label{sec:DAB-Intro}

This section summarizes the version of the DAB algorithm introduced in  \cite{Farsad2020} which will be applied to compute the capacity and associated capacity-achieving distribution of the AC-AWGN channel\footnote{Code at \url{https://github.com/CodePizzeria/DAB-for-AWGN-channels.git}}. For this channel, the input and output alphabet is continuous and uncountable, but a unique finite-support capacity-achieving distribution is known to exist \cite{Smith1971TheIC}. As the original DAB was written for discrete output alphabets, modifications were required to enable compatibility with continuous output alphabets. 


Algorithm 1 summarizes the basic steps of DAB for the AC-AWGN channel. Modifications made to accomodate the PC-AWGN channel are described in section \ref{sec:DAB-AWGN-PC-details}. The general structure of the algorithm: Optimizing probability assignment, determining convergence, adding a mass point when necessary, and moving mass points, is identical for both channels.

At the heart of DAB is 
Csiszar's Min-Max Capacity Theorem \cite{CsiszarInformtionTheoryBook1981}, which  states:
\begin{equation}
C = \min_{p(y) \in \{P_Y\}} \max_x D \bigl (p(y|x)\,\|\,p(y) \bigr )\, , \label{eq:Csiszar}
\end{equation}
where $\{P_Y\}$ is the set of distributions on $Y$ induced by a valid input PMF. 
An upper bound on capacity follows directly from \eqref{eq:Csiszar}:  For any valid output distribution on ${\cal Y}$,
\begin{equation}
\label{eq:D-bound}
C \le \max_x D \left ( P_{Y|X=x}\|P_Y \right ) \, .
\end{equation}

In each DAB iteration, Eq. (\ref{eq:D-bound}) is calculated, and if it is lower than the previous lowest upper bound, update the lowest upper bound to the current value. 


The iterative algorithm for the AC-AWGN channel with SNR $s$ starts by initializing the number of mass points $|\mathcal{X}^{(1)}|$ and their locations $\mathcal{X}^{(1)}$ . The capacity-achieving input cardinality $|\mathcal{X}^*|$  is either $|\mathcal{X}^{(1)}|$ or $|\mathcal{X}^{(1)}|+1$, given $\Delta s$ is small enough, as is apparent from Fig. \ref{fig:peakevolution+SNR}. $|\mathcal{X}^{(1)}|$ is set to the $|\mathcal{X}^*|$ that achieved capacity  under $s - \Delta s$. 

For example, one can start with SNR$= 3$ dB for which the capacity-achieving distribution will for certain be two mass points (Fig. \ref{fig:peakevolution+SNR}). During the $k$-th iteration of the algorithm, first $\mathcal{X}^{(k)}$ is used with the Blahut-Arimoto algorithm to maximize MI and find the corresponding maximizing probabilities over the support set $\mathcal{P}^{(k)}$. This provides a lower bound on capacity.  Then, the distribution $p(y)$ induced by $\mathcal{X}^{(k)}$ and $\mathcal{P}^{(k)}$ are used to compute an upper bound on capacity using \eqref{eq:D-bound}. If these bounds are within a specified threshold $\epsilon$, then we have found the capacity and the optimal PMF, otherwise the location of the mass points and/or the number of mass points needs to be updated in steps 3 and 4 of Algorithm \ref{Alg:ROVAr}. 

\begin{algorithm}[t]
	\caption{Dynamic Assignment Blahut Arimoto (DAB)} 
	\label{Alg:ROVAr}
	
    {\bf Initialization:} 
	For the AC-AWGN channel and SNR $s$, $\mathcal{X}^{(1)}$ is the $\mathcal{X}^*$, which achieves capacity for the AC-AWGN channel with the same AC and SNR $s - \Delta s$ and thus $|\mathcal{X}^{(1)}|$ is set to $|\mathcal{X}^*|$.  The mass point locations are arranged in increasing order in the vector $\mathcal{X}^{(1)} = \begin{bmatrix}x_1& x_2&\ldots &x_{|\mathcal{X}^{(1)}|} \end{bmatrix}$.  The capacity-achieving number $|\mathcal{X}^*|$ of mass points should either be $|\mathcal{X}^{(1)}|$ or $|\mathcal{X}^{(1)}|+1$ for sufficiently small SNR step size $\Delta s$.  Select the tolerance $\epsilon$ controlling the accuracy of the final capacity.  We have typically selected $\epsilon = 10^{-5}$, but this may be more precision than necessary for some applications. 
	
	{\bf Iterations:}  
	Determination of the optimal $\mathcal{X}^*$, $\mathcal{P}^*$, and the capacity $C$ (within $\epsilon$ bits) proceeds as follows:
	\begin{enumerate}
	\item \textit{Finding optimal $\mathcal{P}^{(k)}$: }Given $\mathcal{X}^{(k)}$, use Blahut Arimoto to compute the MI-maximizing PMF $\mathcal{P}^{(k)}$ and the corresponding MI $I^{(k)}$, which is a lower bound on $C$.
	\item \textit{Determining Convergence: }
	\begin{enumerate}
	    \item Use the distribution $p(y)$ induced by $\mathcal{X}^{(k)}$ and $\mathcal{P}^{(k)}$ to compute the capacity upper bound $$D_{\text{max}}^{(k)} = \max_{x \in {\cal X}} D \bigl (p(y|x)\,\|\,p(y) \bigr )$$
    	\item If $D_{\text{max}}^{(k)} -I^{(k)} < \epsilon$ conclude by reporting  $\mathcal{X}^*=\mathcal{X}^{(k)}$, $\mathcal{P}^*=\mathcal{P}^{(k)}$, and $C= I^{(k)}$. Otherwise continue.
	\end{enumerate}
	\item \textit{Adding mass point: }Determine whether $|\mathcal{X}^{(k+1)}| = |\mathcal{X}^{(k)}|$ or $|\mathcal{X}^{(k)}|+1$ and if the latter, update $\mathcal{X}^{(k+1)}$ to include the additional location.
	\item \textit{Improving $\mathcal{X}$: }
	\begin{enumerate}
	    \item Determine direction vector $\mathcal{\tilde{D}}^{k}$ to adjust $\mathcal{X}$.
    	\item Compute 
    	\begin{equation*}
    	\label{eq:X+lambdaD}
    	    \mathcal{X}^{(k+1)} = \mathcal{X}^{(k)}+ \lambda^* \mathcal{\tilde{D}}^{k} \, ,
    	\end{equation*}
    	where
    	\begin{equation*}
    	    \lambda^* = \arg \max_{\lambda} I\left(\mathcal{X}^{(k)}+ \lambda \mathcal{\tilde{D}}^{k}, \mathcal{P}^{(k)} \right) \, ,
    	\end{equation*}
    	and $I(\mathcal{X},\mathcal{P})$ is the mutual information that results from an PMF with mass points whose locations are described by the vector $\mathcal{X}$ with corresponding probabilities are described by the vector $\mathcal{P}$.
	\end{enumerate}
	\item $s = s+\Delta s$
	\item Go to 1.
	\end{enumerate}
	\end{algorithm}

\subsection{Details of DAB for the AC-AWGN channel}
\label{sec:DAB-AWGN-AC-details}
This section provides additional details for each of the four steps of Alg. \ref{Alg:ROVAr}. 
\subsubsection{Finding optimal \texorpdfstring{$\mathcal{P}^{(k)}$}{Probabilities}} 
The Blahut Arimoto Algorithm as described in \cite{Blahut1972} is used to find the optimal probabilities $\mathcal{P}^{(k)}$ over the current support set $\mathcal{X}^{(k)}$.

\subsubsection{Determining Convergence}
All details are in Sec. \ref{sec:DAB-Intro}. 

\subsubsection{Adding Mass Point}
\label{sec:splitting}
Based on the point moving procedure in step 4, an additional mass point is added whenever none of the mass points in $\mathcal{X}^{(k)}$ lies in the interval bounded by $x_{\text{max}}$ and $1/2$, because this indicates that moving any of the points will not decrease the upper bound further.  

If $|\mathcal{X}^{(k)}|$ is even, a new point is added at $1/2$.  If  $|\mathcal{X}^{(k)}|$ is odd, a new point is added by splitting the point at $1/2$ into two points that will then be pulled away from $1/2$ by the line search of step 4b.

\subsubsection{Improving \texorpdfstring{$\mathcal{X}$}{X}}
\label{sec:ImprovingX}

First, a direction $\mathcal{\tilde{D}}^{k}$ is selected along which $\mathcal{X}^k$ will be varied in step 4b to increase the mutual information $I(\mathcal{X},\mathcal{P})$. 

The technique considered in the original DAB  \cite{Farsad2020} was to select and move a single mass point so that $\mathcal{\tilde{D}}^{k}= e_j$. $e_j$ is the vector with all elements set to zero except the $j^{th}$ element, which is set to one. Since the capacity-achieving distribution is symmetric due to uniqueness \cite{Smith1971TheIC}, $\mathcal{\tilde{D}}^{k}$ is instead selected as a symmetric pair of mass points so that 
\begin{equation}
    \mathcal{\tilde{D}}^{k}= e_j + e_{ |\mathcal{X}^{(k)}| +1-j}.
\end{equation}
so that a symmetric pair of mass point locations are set to 1.

As in the original DAB \cite{Farsad2020}, motivated by reducing the upper bound of \eqref{eq:D-bound}, we set $\mathcal{\tilde{D}}^{k}= e_j$ where the mass point $x_j$ is the point in the interval bounded by $x_{\text{max}}$ and $1/2$ that is closest to $x_{\text{max}}$, where
\begin{align}
x_{\text{max}}^{(k)}  &= \arg \max_{x}  D \bigl (p(y|x)\,\|\,p(y) \bigr ) \,.
\end{align}

A line search routine (such as fminbd in Matlab) then determines the value of $\lambda$ that maximizes $I(\mathcal{X},\mathcal{P})$ in step 4b of Algorithm \ref{Alg:ROVAr}. 


\section{Amplitude Constrained AWGN Channel}
\label{sec:DAB-AWGN-AC}
Subsection \ref{sec:DAB-AWGN-AC-results} presents the capacity-achieving distributions, a comparison between the AC-AWGN channel capacity and the PC-AWGN channel capacity, and the input cardinality for a wide range of SNR.

\subsection{Results for AC-AWGN channel}
\label{sec:DAB-AWGN-AC-results}

\begin{figure}[t]
\centering\includegraphics[width=20pc]{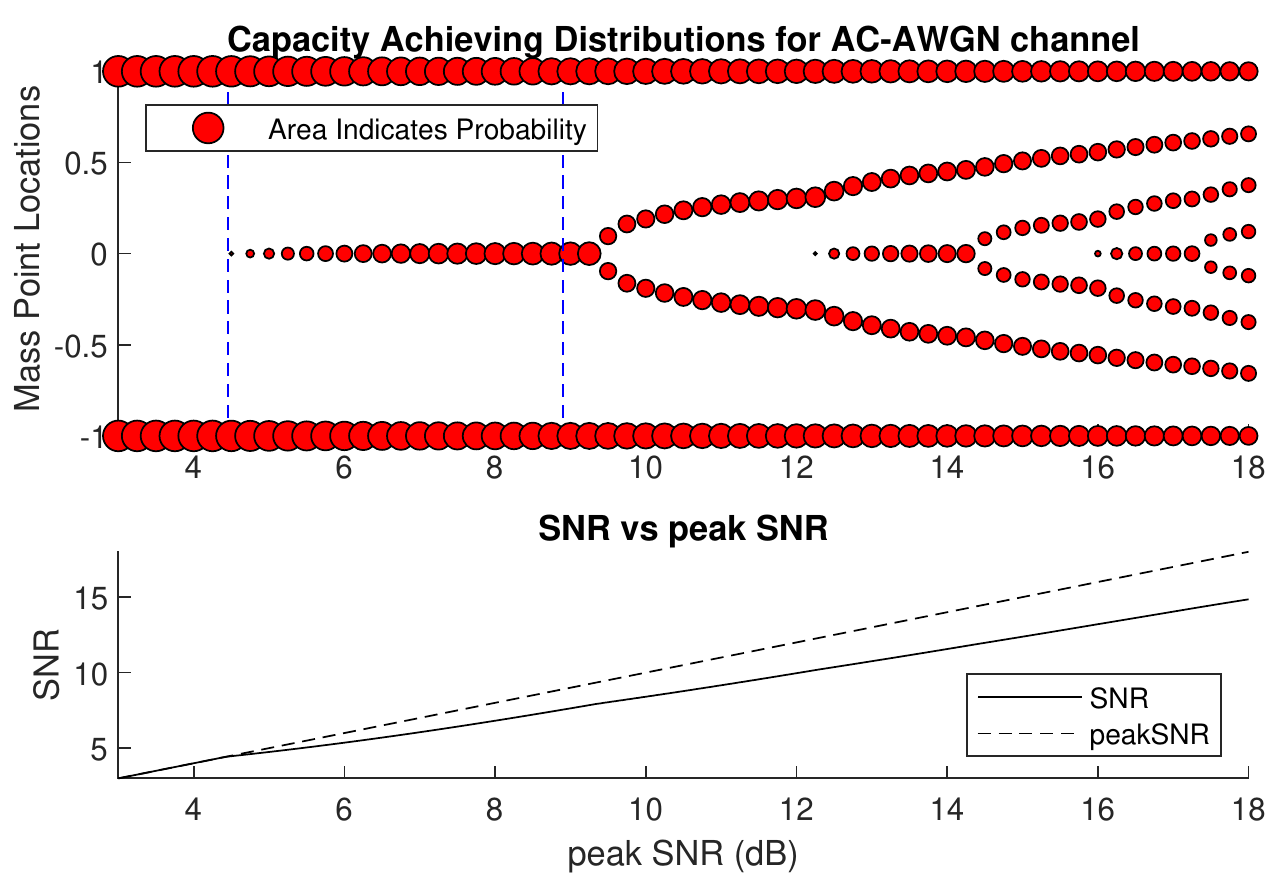}
\caption{(Top) Capacity-achieving input PMFs of the AC-AWGN channel, with at most $10^{-5}$ bit error. The approximate theoretical input cardinality transition points are shown with vertical blue dotted lines. (Bottom) The true SNR of the input is lower than the "peak SNR" defined in Sec. \ref{sec:DAB-AWGN-AC-results}.}
\label{fig:peakevolution+SNR}
\end{figure}

\begin{figure}[t]
\centering\includegraphics[width=20pc]{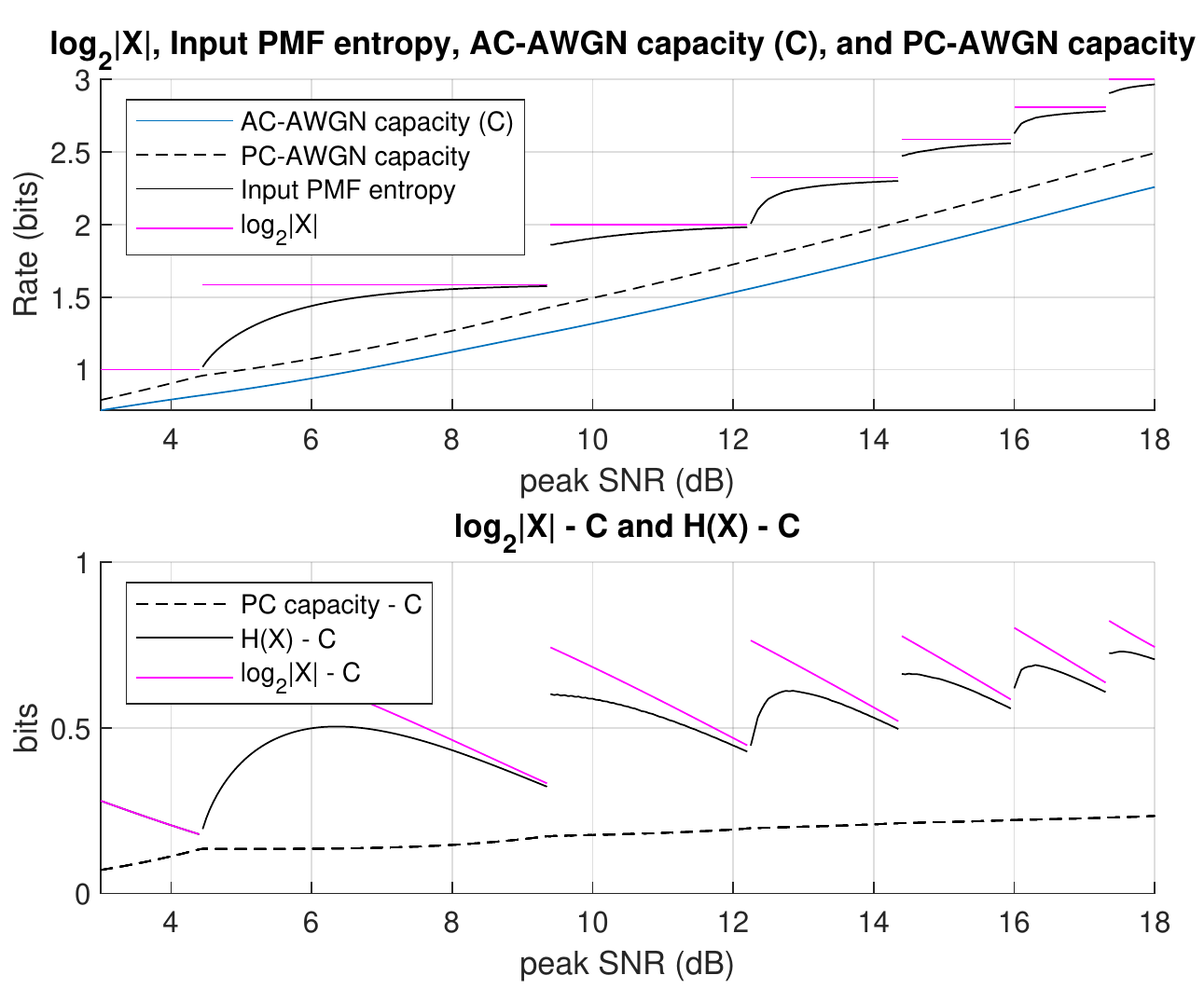}
\caption{(Top) AC-AWGN capacity, PC-AWGN capacity for the power used by the AC-AWGN capacity-achieving input PMF, $\log_2 |{\cal X}|$, and $H(X)$  for the AC-AWGN capacity-achieving input PMF.  (Bottom) Each of the curves in the top figure, with $C$ subtracted, where $C$, in this figure, is the AC-AWGN channel capacity. }
\label{fig:peakcardinality}
\end{figure}

The capacity-achieving inputs PMFs found by DAB achieve capacity with less than an arbitrarily small tolerance $\epsilon$, in this case $10^{-5}$ bits. Theses PMFs are shown in Fig. \ref{fig:peakevolution+SNR}. 

Since the input power is not known before DAB terminates, the distributions are plotted against "peak SNR (dB)" instead of the true SNR, defined as 

\begin{equation}
    \text{peak SNR (dB)}= 10 \log_{10} \frac{1}{N}
\end{equation} where N is the noise power, as if all the probability was assigned to the two peak points. The relationship between peak SNR and SNR is displayed in Fig. \ref{fig:peakevolution+SNR}.

In Fig. \ref{fig:peakcardinality}, We compare the PC-AWGN channel capacity, AC-AWGN channel capacity, and $\log_2$ input cardinality. The PC-AWGN channel capacity is calculated using the true SNR corresponding to each peak SNR. The bottom plot contains the same information, but with the AC-AWGN capacity subtracted. In the bottom plot, $\log_2|\mathcal{X}|$ for the AC-AWGN channel maintains a constant maximum distance from the AC-AWGN channel capacity of around 0.8 bits. Large discontinuous jumps in input entropy correspond to when an odd number of mass points transitions to an even number, i.e. when the mass point at $x=0$ splits in two. Interestingly, the difference between the AC-AWGN capacity and the corresponding the PC-AWGN capacity seems to approach 0.25 bits, which is the shaping loss of equilattice constellations. 

The input cardinality transition points in Fig. \ref{fig:peakevolution+SNR}  where binary and ternary signaling are no longer optimal occur, respectively, at a peak SNR of 4.44 dB and 9.28 dB. The two theoretical transition points approximately evaluated in \cite{Sharma_2008} indicate that for unit variance Gaussian noise, the transition should occur at an amplitude constraint of 1.671 and 2.768, or, in the notation of this paper, a peak SNR of 4.46 and 8.90 dB. Our experimental transition points (to a rate tolerance of $10^{-7}$) differ from those in \cite{Sharma_2008} by 0.02 and 0.38 dB.


\section{AWGN Channel with a Power Constraint}
\label{sec:DAB-AWGN-PC}

The optimization of information rate for the PC-AWGN channel over finite-support (fs) input distributions of a fixed cardinality $N$ can be formulated as 

\begin{equation}
    C_{fs}= \max_{\substack{
                  \mathcal{X}, \mathcal{P} \\
                  \mathbbm{E}(\mathcal{X}^2)\leq E \\
                  |\mathcal{X}| = N
                  }}
                  I\left(\mathcal{X}, \mathcal{P} \right)
\end{equation} 
where $E$ is the signal power constraint, and $\mathcal{X}, \mathcal{P}$ represent the PMF support set and PMF probabilities respectively. 

The PC-AWGN channel capacity without input cardinality constraints is well known to be 
\begin{equation}
    C = \frac{1}{2}\log_2(1+SNR)
\end{equation}
and is achieved by the continuous gaussian PDF. 

Subsection \ref{sec:DAB-AWGN-PC-results} shows that a DAB optimized finite-support input PMF approaches the PC-AWGN channel capacity to within 0.01 bits using an input cardinality of size $2^{C+1.2}$. By comparison. Ungerboeck's rule for equilattice input PMFs \cite{Ungerboeck_1982} establishes that the comparably simpler PMF with input cardinality $2^{C+1}$ achieves only within 0.25 bits of capacity.

Subsection \ref{sec:DAB-AWGN-PC-details} summarizes the modifications required so that the DAB algorithm for the AC-AWGN channel works for the PC-AWGN channel. 

To obtain the results, one instance of DAB finds the cardinality constrained PC-AWGN capacity, and associated capacity-approaching PMF given the SNR $s$ and input cardinality $|\mathcal{X}|$. The next DAB instance uses the previous optimal input PMF as the initial guess, and find the optimal input PMF at $s+\Delta s$ for the same cardinality constraint. This process is repeated until optimal input PMFs have been found for each SNR the SNR interval of interest. This whole process is then repeated with the a cardinality constraint of $|\mathcal{X}|+1$. 

Note that this is different from the AC-AWGN channel DAB algorithm, where only one "evolution" curve is found over the SNR interval of interest. For the cardinality constrained PC-AWGN channel, an evolution curve for each input cardinality of interest is found, and only then is the minimum cardinality input PMF for which $C-C_{fs} \leq 0.01$ bits  chosen. 

\subsection{Modifications to the AC-AWGN channel DAB Algorithm}
\label{sec:DAB-AWGN-PC-details}
The modifications are motivated by the requirement that the input PMF satisfies a power constraint $E$. Each step in the algorithm, therefore, must search only among PMFs that have an average power of $E$. 

\subsubsection{Blahut-Arimoto with a Power Constraint} 
When finding the optimal probabilities $\mathcal{P}^{(k)}$, the PMF must now also satisfy the specified power constraint $E$. 
Blahut \cite{Blahut1972} showed that the Blahut-Arimoto algorithm can be modified to find the capacity with a power constraint parametrized by $s$. 

However, the signal power $E$ resulting from choosing a $s$ is unknown before the Blahut-Arimoto algorithm terminates, and the derivative of $E(s)$ couldn't be found analytically. Since we do not have an analytical derivative, we use the secant method to find the value of $s$ that approaches the required power constraint within a specified tolerance ($10^{-8}$). 

\subsubsection{Determining Convergence}
The relative entropy upper bound used in the DAB algorithm for the AC-AWGN channel is no longer a tight upper bound in this case, so it cannot be used to determine convergence for the PC-AWGN channel. Convergence is instead defined in terms of rate improvement per iteration, where the algorithm terminates if either of these two conditions are satisfied: 

\begin{itemize}
  \item The mutual information increase per iteration drops below a predefined tolerance. We used $10^{-5}$. 
  \item The above condition isn't met by 400 iterations.
\end{itemize}

\subsubsection{Adding Mass Point}
The PC-AWGN capacity-approaching finite-support distributions do not evolve smoothly as they do for the AC-AWGN channel, so this step is removed. The method described in Sec. \ref{sec:DAB-AWGN-PC} is used to generate Fig. \ref{fig:cardinality}.

\subsubsection{Determining direction vector \texorpdfstring{$\mathcal{\tilde{D}}^{k}$}{}}
\label{sec:direction}
In this step of DAB for the AC-AWGN channel (step 4 of Alg. 1), a direction $\mathcal{\tilde{D}}^k$ is selected along which $\mathcal{X}^k$ will be varied in step 4 to increase the mutual information $I(\mathcal{X},\mathcal{P})$. For the PC-AWGN channel, moving a mass point without adjusting the probability assignment violates the power constraint. 

A point pair moving method that adjusts the probabilities to maintain the power constraint is outlined below. Only the case where there are outer mass points with higher power and probability is flowed to or from those outer points is derived. For the outermost points, another method that adjusts the probability on inner points with lower power is used. Due to the similarity, that derivation is omitted. 

Let $p_i$ and $x_i$ be the $i^{th}$ element of  $\mathcal{P}$ and $\mathcal{X}$ respectively, where $\mathcal{P}$ and $\mathcal{X}$ describes the initial finite-support PMF with cardinality $N$. Let the points of interest be the symmetric $j^{th}$ and $(N-j+1)^{th}$mass points. We can further simplfy the equations by grouping the probabilities and powers into outer points, the two points being moved, and inner points:

\begin{equation}
\label{eq:convenient_probability_variables}
  \begin{aligned}
    p_{out} &= p_{1:j-1} + p_{N-j+2:N}\\
    p_{move} &= p_j+p_{N-j+1}\\
    p_{in} &= p_{j+1:N-j}
  \end{aligned}
\end{equation}

\begin{equation}
\label{eq:convenient_power_variables}
  \begin{aligned}
    E_{out} &   = \sum_{i=1}^{j-1} p_ix_i^2 +\sum_{i=N-j+2}^{N} p_ix_i^2 \\
    E_{move} &  = (p_j+p_{N-j+1})x_j^2 \\
    E_{in} &    = \sum_{i=j+1}^{N-j} p_ix_i^2
  \end{aligned}
\end{equation}

The new PMF which adjusts for point moving is described by $p_i'$ and $x_i'$.
When moving points, we maintain the power constraint $E$ by flowing probability to or from the outer points. The equations to be solved are below. The primed $E$ variables are defined as in \eqref{eq:convenient_power_variables} in terms of the new $p_i'$. 
\begin{equation}
\begin{aligned}
p_i' = \begin{cases} 
      \alpha_{out}p_i   & \text{for outer points} \\
      \alpha_{move}p_j  & \text{for points of interest}\\
      p_i               & \text{for inner points}
   \end{cases}
   \end{aligned}
\end{equation}

\begin{equation}
    \begin{aligned}
        p_{out}' + p_{move}' + p_{in}' =& 1\\
        E_{out}' + E_{move}' + E_{in}' =& E
    \end{aligned}
\end{equation}

For step 4a) of Alg. \ref{Alg:ROVAr}, select the point pairs in a round robin fashion. 

For step 4b), the approach is similar, but the maximization is done over adjusted PMFs that satisfy the power constraint.

\subsection{Results for AWGN with a power constraint}
\label{sec:DAB-AWGN-PC-results}
\begin{figure}[t]
\centering\includegraphics[width=20pc]{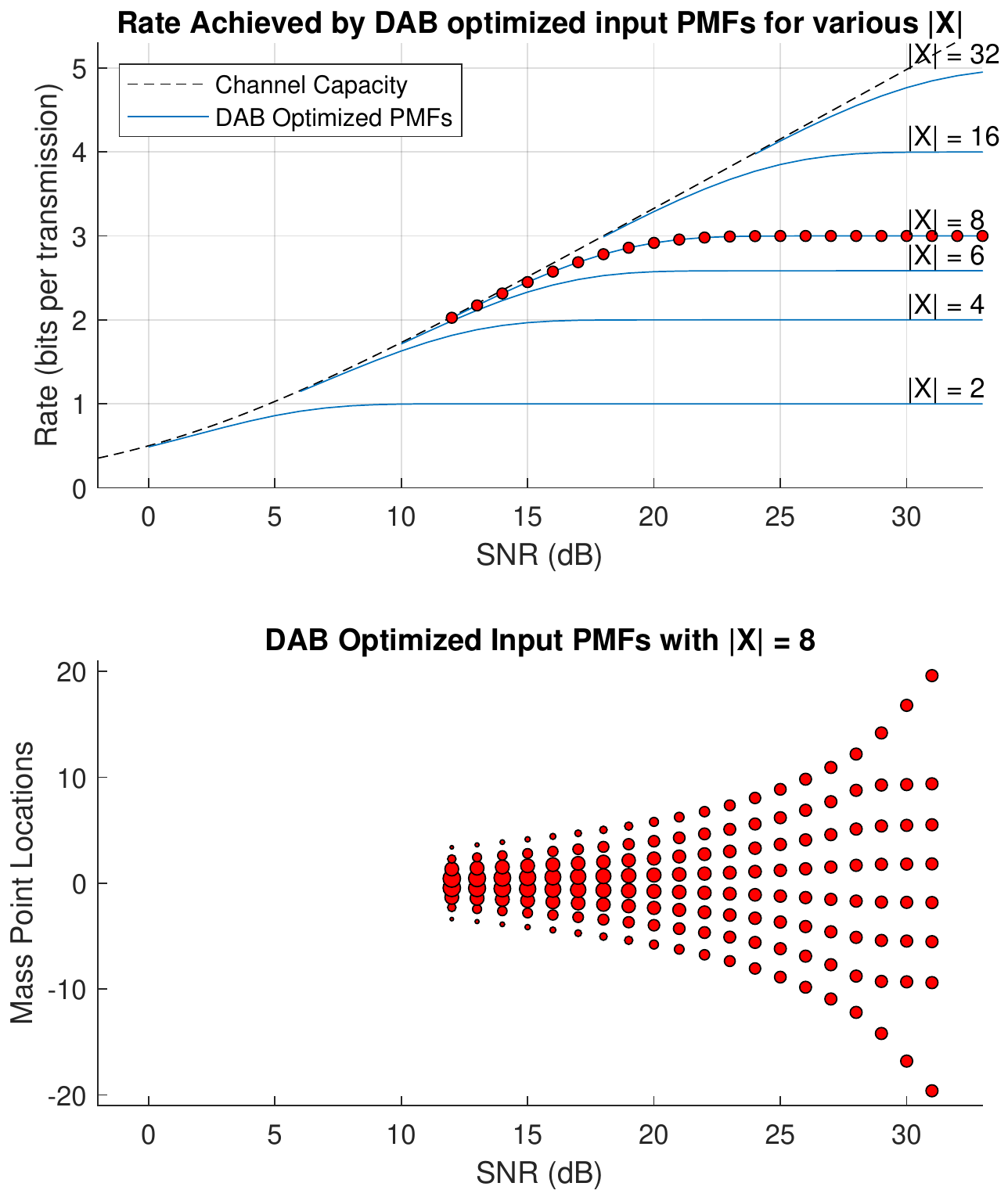}
\caption{(Top) Information rates for the DAB optimized input PMFs for $|\mathcal{X}|$ = 2, 4, 8, 16, and 32, for the PC-AWGN channel. (Bottom) The DAB-optimized capacity-approaching input PMFs for $|\mathcal{X}|$ = 8.}
\label{fig:avg_rate}
\end{figure}
    
The DAB algorithm is used to compute optimal finite-support PMFs for all input cardinalities between 2 and 32, and SNR ranging from 0 to 33. Then, from these PMFs, we select minimum cardinality distributions which are at most 0.01 bits from the continuous PC-AWGN channel capacity. 

The rate achieved by these PMFs are plotted for $|\mathcal{X}|$ = 2, 4, 6, 8, 16, and 32 in Figure \ref{fig:avg_rate}. The curves for the rate achieved by DAB optimized PMFs start where there is a 0.01 bit gap from capacity to reduce visual clutter; most curves approach capacity down to a 0.0001 bit gap. The top figure suggests that a sufficiently larger $|\mathcal{X}|$ can achieve any specified gap to the continuous channel capacity.

The bottom plot in Figure \ref{fig:avg_rate} shows the DAB optimized PMFs for $|\mathcal{X}|=8$. Looking at the top plot, as the rate curve for $|\mathcal{X}|=8$ peels away from the AC-AWGN channel capacity, the optimal PMF shifts from being Gaussian-like to an equilattice-like distribution. This is empirical evidence in support of Wu's proof \cite{Wu2010} that, for fixed $|\mathcal{X}|$, in the low SNR limit, the Gauss quadrature, a gaussian-like construction, converges to the PC-AWGN channel capacity, and in the high SNR limit, the equilattice in \cite{Ungerboeck_1982} converges to the finite-support PC-AWGN channel capacity.  At $SNR = 29$, the rate achieved is 2.999926 bits, leaving at most $7.4*10^{-5}$ bits to improve. Consequently, DAB converges within one iteration, only moving the outermost support points before converging. While this is visually different from the equilattice, it is practically indistinguishable in terms of achieved rate.

\begin{figure}[t]
\centering\includegraphics[width=20pc]{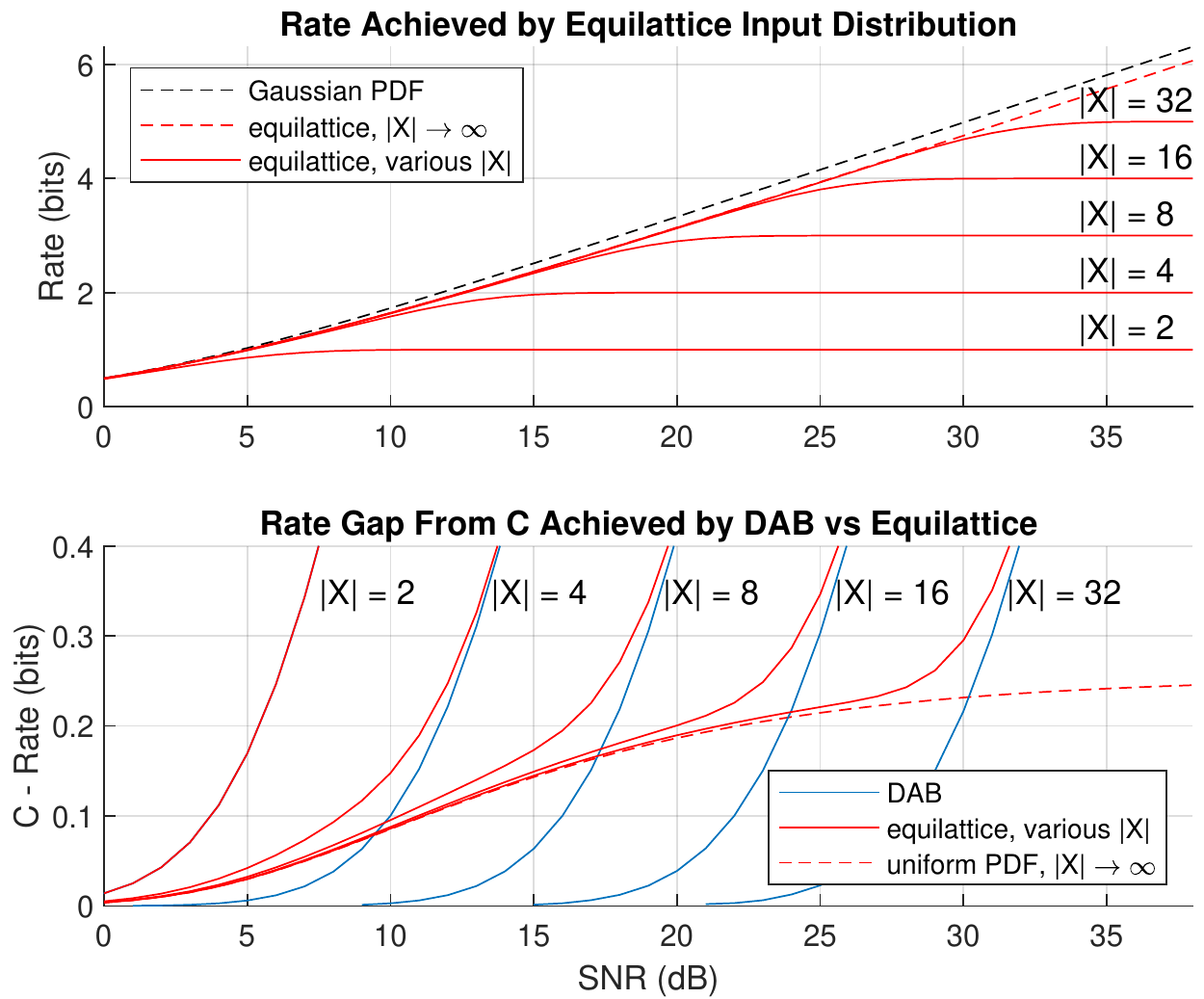}
\caption{(Top) Information rates for the equilattice distributions described by Ungerboeck \cite{Ungerboeck_1982}. (Bottom) Comparison of the information rate gap to PC-AWGN channel capacity achieved by DAB and equilattice. The DAB optimized input distributions consistently outperform the equilattice by using an asymmetric input distribution.}
\label{fig:ungerboeck}
\end{figure}

\begin{figure}[t]
\centering\includegraphics[width=20pc]{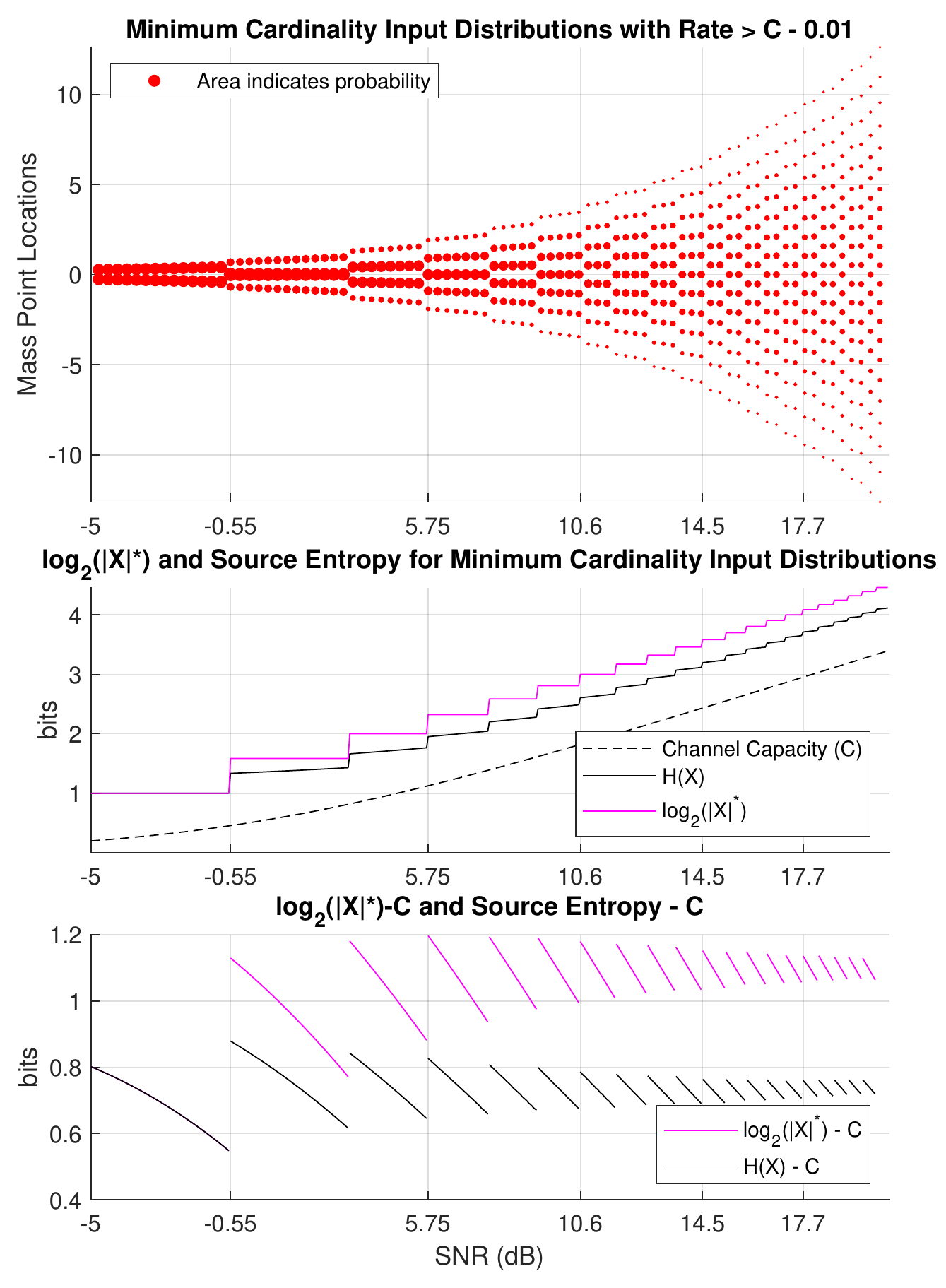}
\caption{(Top) Minimum cardinality capacity-approaching input PMFs (within 0.01 bits of PC-AWGN channel capacity). (Middle) Input cardinality and input entropy compared to the PC-AWGN channel capacity. (Bottom) The middle plot, subtracting the PC-AWGN channel capacity. This demonstrates a rule of thumb for the input cardinality needed to achieve a 0.01 bit gap to channel capacity: "add 1.2 bits to channel capacity"}
\label{fig:cardinality}
\end{figure}

A well-known scheme for generating capacity-approaching finite-support PMFs is the equilattice described in Ungerboeck's paper \cite{Ungerboeck_1982}. The equilattice is a PMF with equiprobable equally spaced mass points with the maximum spacing which satisfies the power constraint.

A replication of the rate curves for the equilattice is shown in the top plot of Fig. \ref{fig:ungerboeck}. The shaping loss of 0.25 bits, seen as a constant width gap between the two dashed lines, cannot be overcome regardless of $|\mathcal{X}|$. The bottom plot shows the rate gap between the continuous PC-AWGN channel capacity and the rate achieved by the DAB optimized PMFs and the equilattice. The shaping loss of the equilattice is clear when looking at the red dotted line. In comparison, the rates achieved by DAB creep right up to the the continuous PC-AWGN channel capacity. 

The relationship between SNR and the minimum cardinality required to achieved a specified rate gap is explored in in Fig. \ref{fig:cardinality}. The top plot shows the  minimum cardinality PMFs which achieve less than a 0.01 bit rate gap. In the middle plot, $\log_2|\mathcal{X}|$ and source entropy are shown, and seem to track channel capacity with a gap that converges to a value. In the bottom plot, the channel capacity is subtracted from  $\log_2|\mathcal{X}|$ and source entropy to elucidate the behavior of the gap. We see that $\log_2|\mathcal{X}|-C$ is upper bounded by 1.2 bits. This gives rise to an analogue of Ungerboeck's well known "add one bit" rule, except our rule is "add 1.2 bits" to achieve within 0.01 bits of capacity by using an asymmetric input PMF.

\section{Conclusion}
\label{sec:Conclusion}

This paper applies the Distributed Allocation Blahut Arimoto (DAB) algorithm to the problem of identifying capacity-achieving finite-support PMFs for amplitude-constrained AWGN channels, where the capacity-achieving distribution is proven to be finite-support and unique  \cite{Smith1971TheIC}. We provide examples of these finite-support capacity-achieving PMFs for various SNR and observe that the optimal input cardinality $|\mathcal{X}|$ is close to the highest $|\mathcal{X}|$ for which $\log_2|\mathcal{X}| - C \leq 0.8$.

As its main contribution, this paper applies the DAB algorithm to the problem of identifying capacity-approaching finite-support PMFs for power-constrained Additive White Gaussian Noise Channels (PC-AWGN channels). DAB ensures that the finite-support PMFs it identifies are within a specified negligibly small distance (0.01 bits) from the PC-AWGN channel capacity, which is theoretically only achieved by a continuous Gaussian pdf. These input PMFs have a cardinality of less than $2^{C+1.2}$, where C is the PC-AWGN capacity. Notably, these input PMFs are of similar cardinality to Ungerboeck's equilattices ($2^{C+1}$) which suffer a far greater shaping loss of  0.25 bits.

\bibliographystyle{IEEEtran}
\bibliography{IEEEabrv,referencesBCH,referencesTBCCvPolar,referencesGlobecom2020DAB,BinomialChannel,literature}

\end{document}